\documentclass[12pt, a4paper]{article}
\usepackage{cite}
\usepackage{amsmath,amssymb}
\input{colordvi.tex}
\usepackage[dvipdfmx]{color}
\usepackage{graphicx}
\usepackage{comment}
\graphicspath{{figures/}}

\newcommand{\eV}{\mathrm{eV}}

\newcommand{\MeV}{\mathrm{MeV}}
\newcommand{\GeV}{\mathrm{GeV}}

\newcommand{\eff}{\textrm{eff}}

\begin{document}

%%%%%%%%%%%%%%%%%%%%%%%%%%%%%%%%%%%%%%%%%%%%%%%%%%

%\tableofcontents

\renewcommand{\thepage}{\arabic{page}}
\setcounter{page}{1}
\renewcommand{\thefootnote}{\#\arabic{footnote}}
\setcounter{footnote}{0}
%%%%%%%%%%%%%%%%%%%%%%%%%%%%%%%%%%%%%%%%%%%%%%%%%%
\begin{titlepage}
	
\begin{center}
		
	\hfill UT-18-20\\
		
	\vskip .75in
		
	{\Large \bf Stochastic Gravitational Waves from Particle Origin}
		
	\vskip .75in
		
	{\large  Kazunori Nakayama$^{(a,b)}$ and Yong Tang$^{(a)}$ }
		
	\vskip 0.25in

	$^{(a)}${\em Department of Physics, Faculty of Science,\\
			The University of Tokyo,  Bunkyo-ku, Tokyo 113-0033, Japan}\\[.3em]
	$^{(b)}${\em Kavli Institute for the Physics and Mathematics of the Universe (WPI),\\
			The University of Tokyo,  Kashiwa, Chiba 277-8583, Japan}
		
\end{center}
	\vskip .5in
	
\begin{abstract}
		
	We propose that there may be a substantial stochastic gravitational wave (GW) background from particle origin, mainly from the gravitational three-body decay of the inflaton. The emitted gravitons could constitute a sizable contribution to dark radiation if the mass of inflaton is close to the Planck scale, which may be probed by future CMB experiments that have a sensitivity on the deviation of the effective number of neutrinos in the standard cosmology, $\delta N_{\textrm{eff}}\sim 0.02 - 0.03$. We have also illustrated the spectrum of the radiated gravitational waves, in comparison to the current and future experiments, and found that GWs from particle origin could be the dominant contribution to the energy density at high-frequency domain, but beyond the sensitivity regions of various GW experiments in the near future. 
		
\end{abstract}
	
\end{titlepage}

\section{Introduction}

Gravitational wave is a general prediction of Einstein's general relativity. Since gravitational interaction is universal and very weak, it can be used to probe directly the very early universe and astrophysical phenomena that happened far away. In the literature, cosmological and astrophysical sources for gravitational waves are studied well. Cosmological sources include primordial vacuum tensor fluctuation during inflation era, possible first-order phase transitions and cosmic strings in particle physics scenarios beyond the standard model~\cite{Binetruy:2012ze}. Astrophysical sources include of stellar collapse and binaries of compact objects~\cite{Maggiore:2018sht}. Except for vacuum fluctuation, the production of gravitational waves from the above sources is determined classically. 

If viewed as quanta like photons, gravitons can be treated as quantum fluctuations over the classical background. Then, it is natural to ask the question that how to generate gravitons or gravitational waves quantum mechanically. Actually, the emission of soft gravitons in elementary processes has been investigated~\cite{Feynman:1963, Weinberg:1965nx}. These results, valid up to a very small momentum or energy, can be applied in a static universe since the wave number could go infinitely small. In an expanding universe, however, proper cut on the low energy has to be imposed as we shall show in this paper. Moreover, if we'd like to go beyond soft limit, dedicated calculations have to be undertaken, especially for high energy gravitons. In Ref.~\cite{Allen:1996vm}, a gravitational wave background with the thermal spectrum, similar to cosmic microwave background (CMB) but with a smaller temperature, was discussed in the context of detection. However, gravitons cannot be in thermal equilibrium after inflation since current CMB experiments have already put constraints on the inflation scale, $\lesssim 10^{16}\,\GeV$, which limits the temperature after inflation should be less than $\sim 10^{15}\,\GeV$. Therefore, gravitons are not in equilibrium after inflation. 

In this paper, we investigate the stochastic gravitational wave from particle decay and show that there could exist a sizable background from the inflaton decay. The spectrum of the resulting gravitational waves approaches $dN/dE\propto 1/E$ at low energy, dramatically different from the thermal spectrum, which can be seen from one of our main results, Eqs.~\ref{eq:scalar} and \ref{eq:fermion}. We also take the cosmic expansion into account and compare the spectrum at present with various ongoing or proposed experiments in Fig.~\ref{fig:gws}, and find that it would be difficult to directly probe such kind of gravitational waves from particle decay.  Production from some particles other than inflaton {\it during} inflation was discussed in~\cite{Senatore:2011sp} which adopted similar formalism to our discussions on soft emissions in Sec~\ref{sec:inflaton}. However, there are crucial differences in setup, spectrum shape and range. The authors in~\cite{Senatore:2011sp} focused on the B-mode detection of primordial gravitational waves produced by an unstable particle which itself was produced {\it during} inflation due to its coupling to inflaton. The wavelength of their interested range therefore can be inflated to a cosmological scale. We, on the other hand, discuss the very short wavelength or very high-frequency regime  which directly results from inflaton decay {\it after} inflation. 

The paper is organized as follows. In Sec.~\ref{sec:formalism} we establish the general framework and formalism for our discussions, and calculate the differential spectrum of gravitons from gravitational three-body decay. Later in Sec.~\ref{sec:inflaton} we apply our results in the inflationary cosmology and show how much relic gravitational wave can be produced from inflaton decay. It will be shown how future CMB would constrain the mass of inflaton through the limit on dark radiation. Also, we compare the predicted spectrum of stochastic gravitational wave from particle decay with the existing experiments. Finally, we give our summary and discussions.

\section{Formalism for Gravitational Decay}\label{sec:formalism}
%%%%%%%%
\begin{figure}[t]
	\begin{center}
		\includegraphics[width=0.85\textwidth,height=0.2\textwidth]{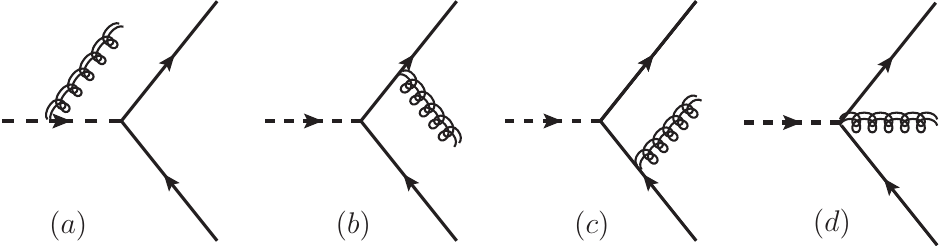}
		\caption{Decay process of a scalar (dashed line) into other particles, involving one graviton (double wiggle line) in the final states. We consider the decay into particles with mass $m$ and spin $s=0,\frac{1}{2}$. The last diagram (d) vanishes for on-shell graviton in our calculation. \label{fig:decay}}
	\end{center}
\end{figure}
%%%%%%%%
We shall explore with the following action in Einstein frame for describing the interaction of a general scalar field $\sigma$ with gravity and other fields, 
\begin{equation}\label{eq:lag}
S=\int d^4x \sqrt{|g|}\left[\frac{M^2_P}{2}R+\frac{1}{2}g^{\mu\nu}\partial_\mu \sigma \partial_{\nu} \sigma - V(\sigma) + \delta \mathcal{L}\right],
\end{equation}
where the reduced Planck scale $M_P=1/\sqrt{8\pi G}\simeq 2.44\times 10^{18}\,\GeV$, $V(\sigma)$ is the potential and $\delta \mathcal{L}$ accounts for other fields and various interactions. For definiteness, we consider the case that $\sigma$ field couples to a canonical scalar $\varphi$ or fermion $\psi$ as
\begin{equation}
\delta L \supset \mu_\varphi \sigma \varphi^\dagger \varphi + y_\psi \sigma \bar{\psi} \psi,
\end{equation} 
where $\mu_\varphi$ and $ y_\psi $ are some couplings that characterize the interaction strength.

%Although we focus on the elaboration of inflation, we shall keep in mind that the following discussions not only apply to inflaton but also other decaying particles as long as they constitute sizable amount of energy density in the early universe. 

To discuss elementary processes with graviton emission, we rely on the following effective interaction
\begin{equation}
	\delta L \supset \frac{\kappa}{2}h_{\mu\nu}T^{\mu\nu},
\end{equation}
where $\kappa\equiv \sqrt{16\pi G}= \sqrt{2}/M_P$, $h_{\mu\nu}$ is the graviton field that defines the quantum fluctuation over the background, $g_{\mu\nu}=\bar{g}_{\mu\nu}+\kappa h_{\mu\nu}$, and $T^{\mu\nu}$ is the energy-momentum tensor of other fields, $\varphi$ and $\psi$. For local particle scattering or decaying processes, we can take a flat spacetime background, $\bar{g}_{\mu\nu}=\eta_{\mu\nu}$, then we can calculate the two-body decay widths for $\sigma$,
\begin{align}
\Gamma_0 (\sigma \rightarrow \varphi ^\dagger + \varphi ) &= \frac{M}{16\pi }\left(\frac{\mu_\varphi}{M}\right)^2\left(1-4y^2\right)^\frac{1}{2}, \\
\Gamma_0 (\sigma \rightarrow \bar{\psi} + \psi ) &= \frac{ y_\psi^2 M}{8\pi }\left(1-4y^2\right)^\frac{3}{2}.
\end{align}
where $m$ is the mass of $\varphi$ or $\psi$, $M$ for $\sigma$ and $y=m/M$.

There are inevitable gravitational decay processes arising from the interactions in Eq.~\ref{eq:lag}. Feynman diagrams for three-body decay, or the graviton bremsstrahlung, processes of the scalar $\sigma$ are shown in Fig.~\ref{fig:decay} where the double wiggle lines represent the gravitons. %Naively, we would expect this kind of gravitational decay has negligible effects to the cosmic relic of gravitons since the branching ratio is suppressed by $M^2_P$. The actual situation is more involved due to the intrinsic properties of gravitational interaction. 
The decay width of such a process is suppressed by $M^2/M^2_P$ as long as the above formalism of effective field theory still holds when $M<M_P$. 
%Accurately speaking, the validity of the perturbative treatment can hold if $\dfrac{\alpha_G}{\pi} < 1$ where $\alpha_G\equiv \dfrac{1}{4\pi}\dfrac{M^2}{M^2_P}$. 
We note that when $M$ approaches $M_P$, our treatment of Einstein's gravity as an effective field theory is subject to theoretical uncertainty due to its nonrenormalizablity since Einstein's gravity might be replaced with some ultraviolet complete theory. Therefore, the uncertainty comes from perturbative expansion of $\dfrac{\alpha_G}{\pi}\left(\alpha_G\equiv \dfrac{1}{4\pi}\dfrac{M^2}{M^2_P}\right)$, should not be taken too literally, especially when $M\gtrsim M_P$.
For processes that radiate a massless graviton with energy $E$ from any initial and final states,\footnote{The last diagram (d) in Fig.~\ref{fig:decay} vanishes for on-shell and traceless gravitons in our calculation.} the rate might be enhanced in the low energy region $E\ll M$, which could be good news for gravitational-wave detectors that target low-frequency. This kind of spectral behavior can be derived by calculating the decay diagram in Fig.~\ref{fig:decay}. For example, Fig.~\ref{fig:decay}(a) alone would contribute to the decay width with the following factor
\begin{equation}
\Gamma_1 \simeq  \Gamma_0\times \left(\frac{1}{16\pi^2}\frac{M^2}{M_P^2}\int_{\Lambda}^{M/2} \frac{dE}{E}\right), 
\end{equation}
where $\Gamma_0$ is the decay width without graviton and $\Lambda$ is the low energy threshold below which particle description is not accurate any more. 
It is not trivial to which frequency of the graviton we can use the three-body decay rate of the inflaton as a particle, because of possible coherence of the inflaton background. As a crude estimation, if the wavelength of the graviton is (at least) shorter than the typical separation of the inflaton particle, $(n_\sigma)^{-1/3}$ with $n_\sigma$ being the inflaton number density, our estimation may be valid. Below we just assume that $\Lambda$ is much smaller than $M$.
%We shall take $1/\Lambda$ as the typical separation distance of particle $\sigma$, namely, $\Lambda \simeq (n_\sigma)^{1/3}$. Here $n_\sigma$ is the number density of $\sigma$ in our Universe, which could vary dramatically in the early time.  
%Note that if the factor in the parenthesis is larger than $1$, the perturbative treatment will break down and a resummation is expected to undertake since processes with radiating two and more gravitons are also important. 

%%%%%%%%
\begin{figure}[t]
	\begin{center}
		\includegraphics[scale=0.81]{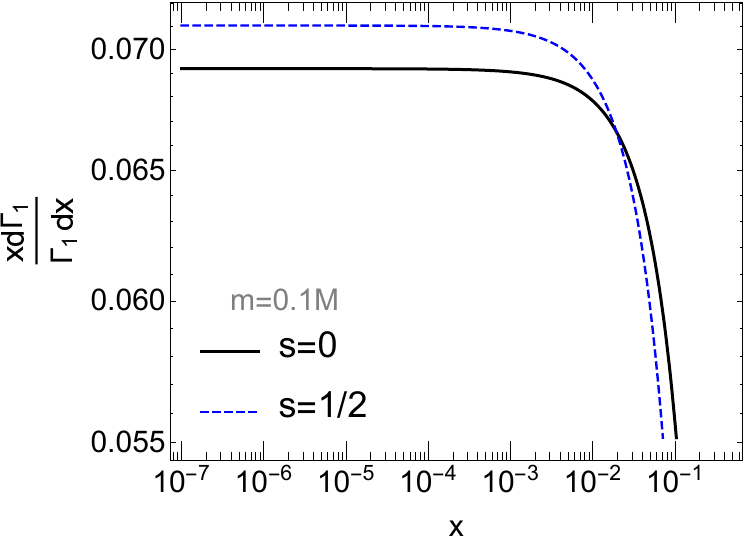}
		\caption{Illustration for normalized spectrum of graviton $\dfrac{xd\Gamma_1}{\Gamma_1dx}$  in three-body decay when $\Lambda=10^{-7}M$. The solid (dashed) curve is for scalar (fermion) case.  \label{fig:hard}}
	\end{center}
\end{figure}
%%%%%%%%

%%%%%%%%
\begin{figure}[t]
	\begin{center}
		\includegraphics[width=0.48\linewidth,height=0.42\linewidth]{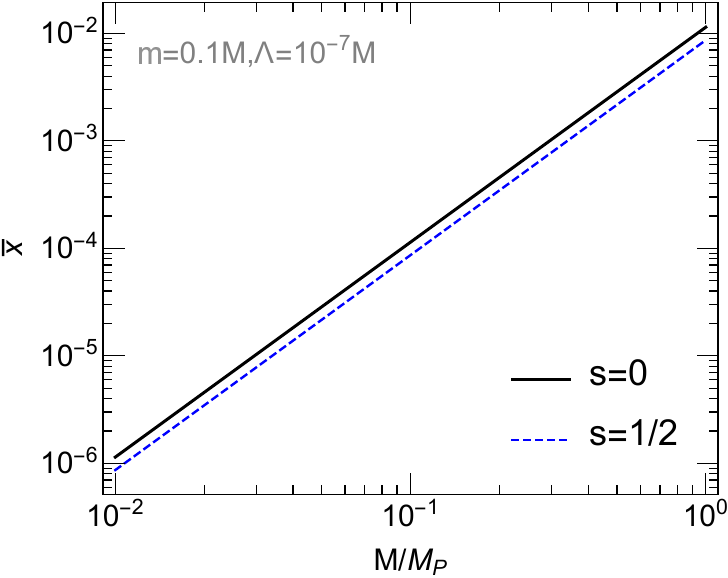}
		\includegraphics[width=0.48\linewidth,height=0.42\linewidth]{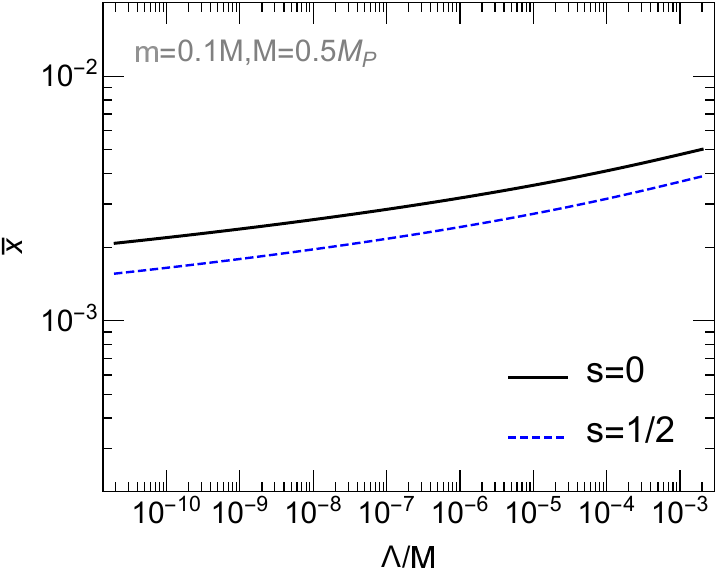}
		\caption{The fraction of energy that goes into gravitons, $\bar{x}$, as a function of $M$ (left panel) and $\Lambda$ (right panel). The solid (dashed) curves are for scalar (fermion) cases.  \label{fig:xML}}
	\end{center}
\end{figure}
%%%%%%%%

After straightforward but tedious calculations, we obtain the full expression for spectrum of the graviton $h$ with energy $E$ in the bosonic decay ($\sigma \rightarrow \varphi ^\dagger + \varphi + h$), which is determined by the differential decay width, 
\begin{align}
\frac{d\Gamma_1}{Mdx} = & \frac{\left(\mu_\varphi/M_p\right)^2}{64\pi^3}\left[\frac{y^2 (-1 + 2 x + y^2)}{x}\ln{\left(\frac{1+\alpha}{1-\alpha} \right)} 
%\nonumber \\ & 
+ \frac{(1 - 2 x) (1 - 2 x + 2 y^2)}{4x\alpha^{-1}}\right],\label{eq:scalar}
\end{align}
where new variables $x=E/M$ and $\alpha=\sqrt{1-\dfrac{4y^2}{1-2x}}$ are defined for convenience. For fermionic decay ($\sigma \rightarrow \bar{\psi} + \psi +h $), the differential decay width is 
\begin{align}
\frac{d\Gamma_1}{Mdx}=  \frac{y_\psi^2\left(M/M_p\right)^2}{64\pi^3}&\left[ -\frac{2y^2 \left[( x-1)^2 + (8 x-5) y^2 + 4 y^4\right]}{x} \ln{\left(\frac{1+\alpha}{1-\alpha}\right)} \right. \nonumber \\
{}\;&  \left. + \frac{(1 - 2 x)\left[1 + (1 - 2 x)^2 + 4 ( 4 x-1) y^2 - 16 y^4\right]}{4x\alpha^{-1}}\right].\label{eq:fermion}
\end{align}
Note that the above formulas~\footnote{They are now consistent with ref.~\cite{Barman:2023ymn}. Although the quantitative results depend on the exact form of these formulas, the essential physical picture does not change. Therefore, we do not update the plots and the rest of discussions are still based on earlier version. } depend on the mass $m$ nontrivially. Since it is only logarithmic dependence when $m\ll M$, for illustrative purposes and simplicity, we shall take $m=0.1M$ in all decay widths throughout our later discussions, if not stated otherwise. 

The typical shape of the normalized spectrum $\dfrac{xd\Gamma_1}{\Gamma_1dx}$ is shown in Fig.~\ref{fig:hard} when choosing $\Lambda=10^{-7}M$, solid (dashed) curve for scalar (fermion). It is evidently seen that the spectrum goes flat in the low energy region, which is expected since the leading contribution in $d\Gamma_1/Mdx$, Eqs.(\ref{eq:scalar},\ref{eq:fermion}), has $1/x$ behavior. It also indicates that  most of the emitted gravitons could be in the lower energy range. 

Although the branching ratio of gravitational three-body decay can be $\mathcal{O}(10\%)$ as $M\rightarrow M_P$, the emitted gravitons only take a small amount of the total energy. As a better illustrative quantity, the total fraction of energy which are injected into gravitons, $\bar{x}$, is determined by 
\begin{equation}
\bar{x} \equiv\frac{\overline{E}}{M}=\frac{\Gamma_1}{\Gamma}\int_{\Lambda/M}^{1/2}\frac{xd\Gamma_1}{\Gamma_1dx} dx, 
\end{equation}
where $\Gamma=\Gamma_0+\Gamma_1$. 
In Fig.~\ref{fig:xML} we show how $\bar{x}$ changes as a function of $M$ (left panel) and $\Lambda$ (right panel). We can see that $\bar{x}$ depends on $M$ quadratically and in general is smaller than $10^{-2}$ for $M<M_P$, and that $\bar{x}$ has not changed much over a large energy range of $\Lambda$.

\section{Application to Inflaton}\label{sec:inflaton}
%%%%%%%%
\begin{figure}[tb]
	\begin{center}
		\includegraphics[width=0.58\linewidth,height=0.45\linewidth]{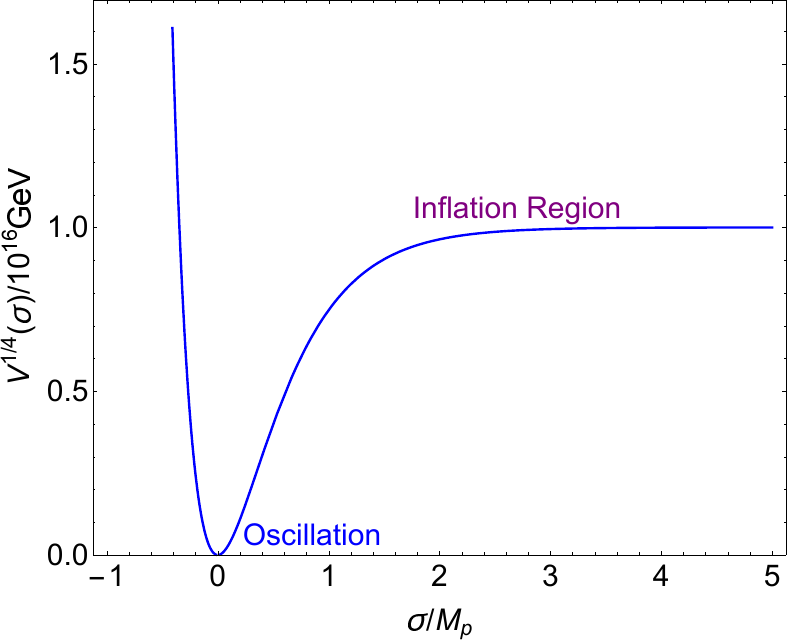}
		\caption{A general potential of an inflation field.\label{fig:Vinf}}
	\end{center}
\end{figure}
%%%%%%%%

Now we are in a position to discuss some applications in an inflationary setup. To keep our discussion as general as possible, we do not specify the inflation model and work in the Einstein frame with action,
\begin{equation}
S=\int d^4x \sqrt{|g|}\left[\frac{M^2_P}{2}R+\frac{1}{2}g^{\mu\nu}\partial_\mu \phi \partial_{\nu} \phi - V(\phi) + \delta \mathcal{L}\right],
\end{equation}
$\phi$ is the inflaton field, $V(\phi)$ is the potential, and $\delta \mathcal{L}$ accounts its various interactions. To describe the perturbative reheating process around the minimum, we expand $V(\phi)$ around $\phi_0$ where $V'(\phi_0)=0$ ,
\begin{equation}
V(\phi)=V(\phi_0)+\frac{1}{2}M^2 (\phi-\phi_0)^2 + \frac{1}{3!}\mu (\phi-\phi_0)^3 + \frac{1}{4!}\lambda (\phi-\phi_0)^4 + ... ,
\end{equation}
where 
\begin{equation}
M^2 = V''(\phi_0),\; \mu=  V'''(\phi_0),\; \lambda=  V''''(\phi_0).
\end{equation}
$V(\phi_0)$ also has to be tiny, less than the current dark energy density. Note that $V(\phi_0)$ is different from $V_{\textrm{inf}}$ which is the typical potential energy during inflation. Defining new field variable $\sigma=\phi-\phi_0$, we then have
\begin{equation}
V(\sigma)\simeq \frac{1}{2}M^2 \sigma ^2 + \frac{1}{3!}\mu \sigma^3 + \frac{1}{4!}\lambda \sigma^4 + ... ,
\end{equation}
where $M$ denotes the inflaton mass throughout our discussion. CMB and LSS observations can probe the dynamics during slow-roll inflation region when the comoving scale close to the present Hubble radius exits the horizon. However, very little is known about the vicinity $\sigma \sim 0$. An intuitive picture is shown in Fig.~\ref{fig:Vinf} where inflation happens in the flat region, while the mass of $\sigma$ is defined when $\sigma$ oscillates around the potential minimum. In principle $M$ could be arbitrary and we shall show gravitational decay is able to constrain the mass $M$, the curvature of inflationary potential at the minimum.\footnote{
	However, the most known high-scale inflation models predict the inflaton mass around $10^{-5} M_P$.
}

During the perturbative reheating era, the universe is in the matter-dominant phase, dominated by the coherent oscillation of the inflaton $\sigma$. 
%The number density is easily obtained $n_\sigma = V/M$ while 
The Hubble parameter is $H=\sqrt{\rho_\sigma/3}/M_P$ with $\rho_\sigma$ denoting the total inflaton energy density and the reheating temperature is $T_R\simeq 0.3 \sqrt{\Gamma M_P}$. In instantaneous reheating approximation, two parameters are enough to characterize the later energy content, which we shall take $H=\Gamma$ at the time of inflaton decay and the mass $M$ as the free parameters for numerical evaluations. 

\subsection{Constraint from $N_\eff$}
The emitted gravitons from the gravitational decay of the inflaton will contribute to dark radiation and change the effective number of neutrinos $N_\eff$ by an amount of
\begin{equation}
\delta N_\eff \simeq \frac{4g_s(T_{E})}{7}\left[\frac{g_s(T_\nu)}{g_s(T_{E})}\right]^{4/3}\frac{\bar{x}}{1-\bar{x}},
\end{equation}
where $g_s(T)$ denotes the total degree of freedom for relativistic SM particles at temperature $T$, $T_\nu$ for neutrino decoupling temperature and $T_E$ for the temperature when SM particles begin to be equilibrated thermally (we shall take $T_E=T_R$ for simplicity, in principle they can be different). We have $g_s(T_\nu)=43/4$, $g_s(T_E)=427/4$ for $T_E>m_t$ ($m_t$ denotes the top quark mass) and  $g_s(T_E)=57/4$ for $m_\mu<T_E<m_\pi$ ($m_\mu$ and $m_\pi$ denote the muon and pion mass, respectively), for instance. Therefore we have
\begin{equation}
\delta N_\eff \simeq \frac{\bar{x}}{1-\bar{x}}\times 
\begin{cases} 
2.86 & m_t< T_s \\
5.59 & m_\mu<T_s<m_\pi
\end{cases}.
\end{equation}
For future CMB experiments that can probe $\delta N_\eff \sim 0.02-0.03$~\cite{Abazajian:2016yjj}, it is possible to constrain $\bar{x}\simeq 10^{-2}$ when $M$ is very close to $M_P$, although the perturbative calculation may not be fully reliable in this limit as noted in Sec.~2.

%%%%%%%%
\begin{figure}[tb]
	\begin{center}
		\includegraphics[width=0.58\linewidth,height=0.45\linewidth]{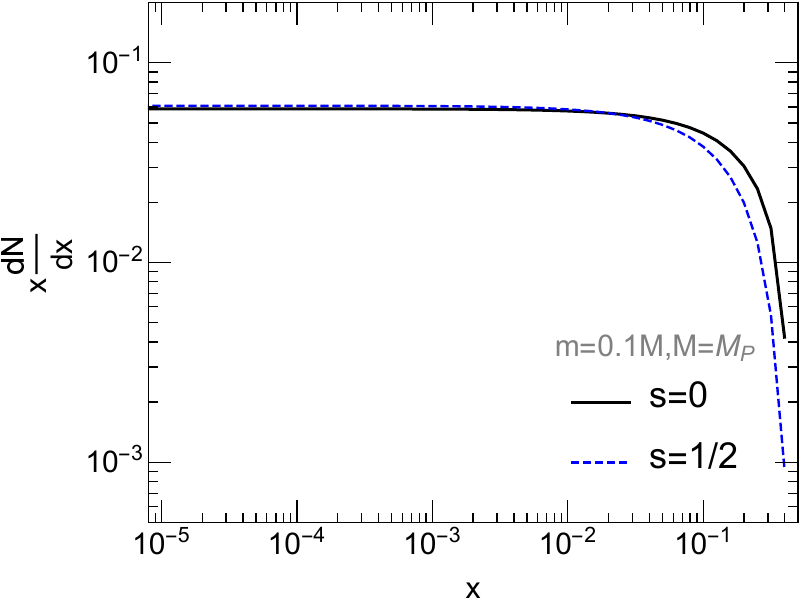}
		\caption{The integrated spectrum for $x\frac{dN}{dx}$ at $t=10^8t_P$ when taking $\Gamma=10^{-8}M_P, M=M_P, \rho^{1/4}_\sigma(z_i)=10^{-4}M_P$.  \label{fig:Int}}
	\end{center}
\end{figure}
%%%%%%%%
In the above estimation, we have used the instantaneous reheating which is actually a good approximation for our interest here. We can show that when taking the finite decay time into account, the results do not change much. For precise evaluation, we have to integrate spectrum over redshift
\begin{align}
	E\frac{dn_h(z)}{dE}=&E\int_{z}^{+\infty}dz'\frac{\Gamma_1 n_\sigma (z') }{(1+z')H(z')}\frac{a^3(z')}{a^3(z)} \frac{dE'}{dE}\left[\frac{dN(E')}{dE'}\right]\\ \nonumber
	=& \dfrac{a^3(z_i) n_{\sigma}( z_i)}{a^3(z)}\int_{z}^{z_i}dz'\frac{\Gamma_1 \exp{[-\Gamma t(z')]} }{(1+z')H(z')}\left[\frac{x'd\Gamma_1}{\Gamma_1dx'}\right]
\end{align}
where $H$ is the Hubble parameter, $z$ the is redshift, $a(z)(1+z)=1$, and $E'=\dfrac{1+z'}{1+z}E$. In our case $dN/dE=d\Gamma_1/\Gamma_1 dE$. We have also used the relation that the number density of inflaton obeys the following equation after inflation
\begin{equation}
n_\sigma (z') = \dfrac{a^3(z_i) n_{\sigma}( z_i)}{a^3(z')}\exp{[-\Gamma t(z')]},
\end{equation}
where we have take time $t=0$ and redshift $z=z_i$ at the point when inflation ends and inflatons are treated as particles. In Fig.~\ref{fig:Int}, we plot the following quantity 
\begin{equation}
x\frac{dN}{dx}(z)=\frac{\Gamma}{\Gamma_1}\int_{z}^{z_i}dz'\frac{\Gamma_1 \exp{[-\Gamma t(z')]} }{(1+z')H(z')}\left[\frac{x'd\Gamma_1}{\Gamma_1dx'}\right],
\end{equation}
at time $t=10^8t_P$ ($t_P$ is the Planck time) by taking an illustrating case with parameters, $\Gamma=10^{-8}M_P, M=M_P, \rho^{1/4}_\sigma(z_i)=10^{-4}M_P$. We observe that the overall shape of the spectrum does not change much from Fig.~\ref{fig:hard}. From now on, we shall use instantaneous reheating throughout our later discussions.

Let us comment on $\Lambda$ in a realistic reheating history. If we assume $\Lambda \sim (n_\sigma)^{1/3}$, $\rho_\sigma=M n_\sigma$ and reheating temperature must be higher than MeV to have a successful big-bang nucleosynthesis (BBN), we would get 
\begin{equation}
M n_\sigma \sim  \MeV^4 \Rightarrow \Lambda \sim \left(\frac{\MeV}{M}\right)^{1/3}\MeV. 
\end{equation}
When $M$ is as high as $M_P$, $\Lambda\sim 10^{-7}$MeV, which is 7 orders of magnitude less than typical photon energy at BBN time. After that, the frequency will get redshifted, suggesting that the current boundary for gravitational waves from particle decay is around $10^4$Hz. Although the frequency could fall into the sensitive range of LIGO experiment, as we shall show immediately, the overall magnitude is way below all existing experiments' probes. 

\subsection{Spectrum of Gravitational Waves at Present}
%%%%%%%%
\begin{figure}[tb]
	\begin{center}
		\includegraphics[width=1\linewidth,height=0.75\linewidth]{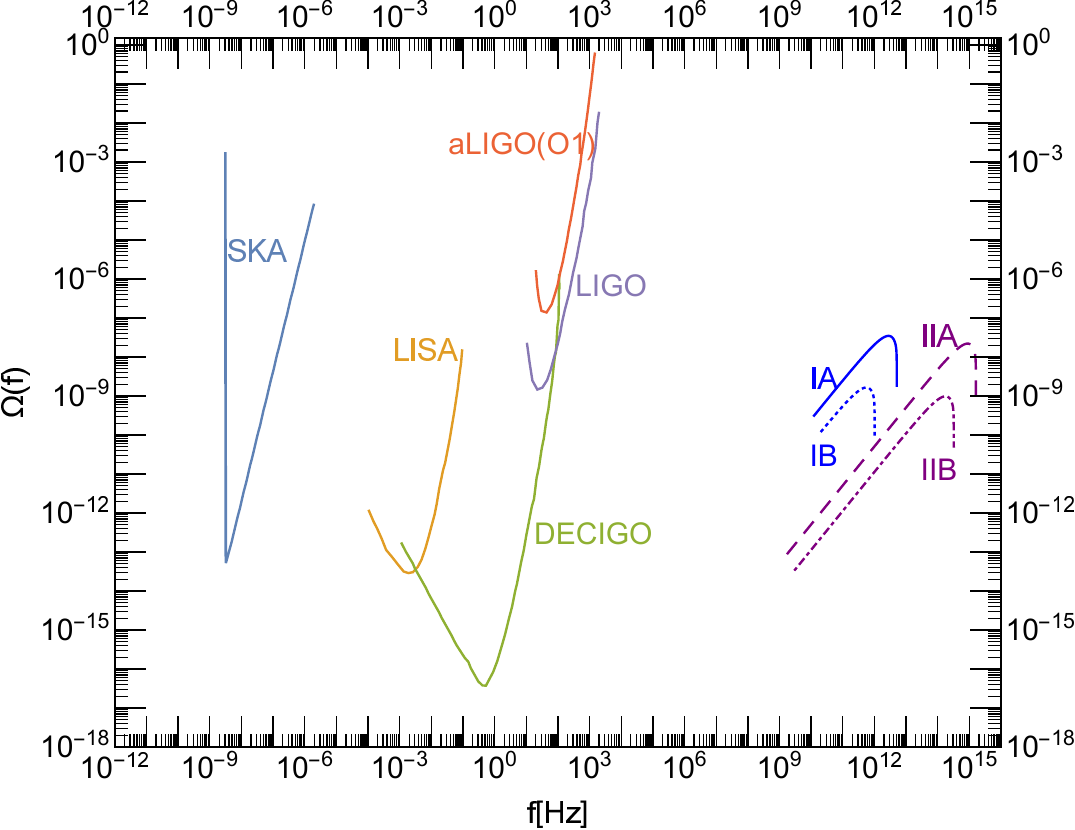}
		\caption{The predicted stochastic gravitational waves from particle decay vs sensitivities of various experiments. See the text for details about the parameters that define the illustrated four cases.  \label{fig:gws}}
	\end{center}
\end{figure}
%%%%%%%%

To compare with sensitivities of various gravitational-wave experiments, we define the following quantity on frequency domain,
\begin{equation}
\Omega(f)\equiv \frac{1}{\rho_c}\frac{d\rho}{d\ln f},
\end{equation}
where the critical energy density $\rho_c=3M^2_P H^2_0$ and the present frequency of gravitational wave $f=E_0/2\pi$. We can rewrite the above equation as
\begin{equation}
\Omega(f)= \frac{\rho_{\gamma }}{\rho_c}\left[\frac{1}{\rho_{\gamma}}\frac{E_0 dn}{d\ln E_0}\right],
\end{equation}
where the present value of $\rho_{\gamma }/\rho_c$ is $5.38 \times 10^{-5}$ and the factor in bracket is approximately constant in cosmic evolution after BBN, therefore it can be evaluated in the early time. Approximating with instantaneous reheating and taking into account decouple of heavy particles, we have   
\begin{equation}\label{eq:omega}
\frac{1}{\rho_{\gamma}}\frac{Edn}{d\ln E} \simeq \left[\frac{2}{g_s(T_{R})}\right]^{1/3}\frac{\Gamma_1/\Gamma}{1-\bar{x}}\frac{x^2d\Gamma_1}{\Gamma_1dx}.
\end{equation}
Note that we should take redshift into account and evaluate the above function at $E\simeq 2\pi f\times T_R/T_{\gamma 0}$ for present frequency $f$ and CMB temperature $T_{\gamma 0}=2.73\,{\rm K}=8.6\times 10^{-5}\,\eV$. 

In Fig.~\ref{fig:gws} we plot the predicted spectrum from bosonic decay (very similar for fermionic decay except at high energy tail), in comparison with current or future gravitational-wave experiments, SKA~\cite{Janssen:2014dka}, LISA~\cite{Audley:2017drz}, DECIGO~\cite{Kawamura:2011zz}, LIGO's first operation (O1) and LIGO designed~\cite{TheLIGOScientific:2016dpb}. There are four illustrating cases defined by the following parameter pairs, 
\begin{itemize}
	\item IA: $M=0.5M_P, \Gamma=10^{-5}M_P$, 
	\item IB: $M=0.1M_P, \Gamma=10^{-5}M_P$,
	\item IIA: $M=0.5M_P, \Gamma=10^{-10}M_P$, 
	\item IIB: $M=0.1M_P, \Gamma=10^{-10}M_P$,
\end{itemize}
%and the Hubble scale $H=\Gamma$, which should be less or equal than scale during inflation.
and the dominant contribution is from those produced at $H = \Gamma$, which should be less than or equal to the Hubble scale during inflation.
As shown in the figure, the peaks of cases I are different from cases II, which is due to the redshift from cosmic expansion. Since they have different Hubble parameters, the reheating temperatures are different, $T_R\simeq 0.3 \sqrt{\Gamma M_P}$, and the peaks are around energy $MT_{\gamma 0}/T_R$. For the same $\Gamma$, larger $M$ would give stronger gravitational emission, therefore the overall height of the spectrum is enhanced, which is proportional to $(M/M_P)^2$. Note that in all cases the spectrum have a finite range due to the energy threshold $\Lambda$ below which particle description is not accurate. It is shown that, although gravitational waves from particle origin could be the dominant contribution to the energy density at high frequency, they are beyond the sensitivity regions of ongoing and future experiments. 

From the above discussion, we would expect that although decreasing $\Gamma$ or $H$ could broaden the frequency domain of spectrum, the overall strength is reduced because $\Omega(f)$ is proportional to $E$ at low energy, which can be easily seen from Eq.~\ref{eq:omega} and $d\Gamma_1/dx \propto 1/x$. In cases where there are some other long-lived particles, such as dark matter, decaying in the present time, the resulting spectrum would be either in very high frequency region or much below the plotted $\Omega(f)$ range.

\subsection{Contribution From Soft Emission}
Although it is not the main contribution, soft graviton emission always exists. The spectrum of soft massless particles (photon and graviton) with $E<T$, where $T$ is temperature of the plasma, has an universal behavior due to the infrared effects~\cite{Weinberg:1965nx}. For minimal coupled matter content, we can show the differential number density for graviton emitted in $\Lambda<E<T$ in a Hubble time is given by
\begin{equation}
\frac{dn}{dE}\approx\frac{n_\sigma F}{1+F}\frac{1}{E}\frac{\mathcal{A}}{H},
\end{equation}
where $\mathcal{A}$ is interacting/scattering rate without graviton, and factor $F$ is the following quantity~\cite{Weinberg:1965nx},
\begin{equation}
F=\frac{1}{16\pi^2 M^2_P}\sum_{ij}\eta_i \eta_j m_i m_j \frac{1+\beta_{ij}^2}{\beta_{ij}\sqrt{1-\beta_{ij}^2}}\ln{\left(\frac{1+\beta_{ij}}{1-\beta_{ij}}\right)},
\end{equation}
where $\eta_i=+1$ or $-1$ for an outgoing or ingoing particles, $m_i$ is the mass of particle $i$, and the relative velocity of $i$ and $j$ in the rest frame of either:
\begin{equation}
\beta_{ij}\equiv \sqrt{1-\frac{m_i^2m_j^2}{\left(p_i\cdot p_j\right)^2}}. 
\end{equation}
Here $p_i$ is the momentum for particle $i$. In cases where there is only one heavy particle $i$ with mass $M$ that dominates, we have $\beta\simeq 0$ and $8\pi^2 F\simeq M^2/M_P^2$. 

There should be an upper bound on the energy of soft graviton, which we shall take as the temperature of inflaton, $T$. For instantaneous reheating, $T$ could be as high as the inflation scale, $T \sim V^{1/4} \sim 10^{15}$\,GeV. If the inflaton lifetime is very long, in principle $T$ could be as low as MeV if the elastic scattering rate is large enough. However, explicit calculation shows that the kinetic equilibrium would require $T\gtrsim M^2/M_P$. Therefore, if $M$ is small enough, inflaton would be in kinetic equilibrium with SM particles, but then the spectrum is highly suppressed because of the factor $F\propto M^2/M_P^2$. Furthermore, the interacting rate $\mathcal{A}$ is generally much smaller than $H$ unless at high temperature, giving an additional suppression. In conclusion, for our purpose, soft emission can be neglected in comparison to the gravitational three-body decay.

\section{Summary and Discussion}\label{sec:summary}
We have investigated the stochastic gravitational waves from particle origin, especially from inflaton decays. We have shown that gravitational decay is unavoidable and the resulting emitted graviton from three-body decay could contribute to dark radiation or the effective number of neutrinos, $\delta N_\eff$. The future CMB experiment would have a sensitivity of $\delta N_\eff \sim 0.02-0.03$, which may probe and constrain an inflaton with mass near the Planck scale. 
We have also calculated the full differential decay width for the gravitational decay of the inflaton, Eqs.~\ref{eq:scalar} and \ref{eq:fermion}, which are crucial for determining the graviton spectrum. One unexpected result is that the emitted gravitons can take at most $\sim 10^{-2}$ of the total energy because the emission rate is stronger for low-energy gravitons. We have also compared the resulting stochastic gravitational wave with the existing experiments in Fig.~\ref{fig:gws}, and found that although constituting as a sizable background, gravitons from particle decay are beyond current experimental sensitivities. To search for such cosmic relics, new innovative ideas need to be pursued to probe high-frequency gravitational waves, although preliminary attempt~\cite{Akutsu:2008qv} have not reach the required sensitivity yet. 

Let us comment on other possible gravitational wave sources with very high frequency.
The so-called gravitational particle production~\cite{Parker:1969au} necessarily leads to the quantum production of gravitons at the end of inflation~\cite{Ford:1986sy,Peebles:1998qn} and also during the inflaton oscillation era~\cite{Ema:2015dka,Ema:2016hlw}.
It is shown in \cite{Ema:2015dka,Ema:2016hlw} that such a graviton production process may be regarded as inflaton ``annihilation'' into the graviton pair whose rate is $\Gamma_{\sigma\sigma\to hh}\sim H^2 M / M_P^2$ and hence the typical momentum of the graviton produced in this way is the inflaton mass. Comparing it with the three-body decay rate $\Gamma_1$ in this paper, we have
\begin{align}
	\frac{\Gamma_1}{\Gamma_{\sigma\sigma\to hh}} \sim \frac{M \Gamma}{H^2}.
\end{align}
At least around $H \sim \Gamma$, the three-body decay contribution is dominant since $M > \Gamma$ should be satisfied for perturbative interactions. At earlier time, it might be possible that the annihilation contribution is dominant depending on parameter choices.

Inflationary gravitational wave background has a flat spectrum toward high frequency~\cite{Turner:1993vb,Turner:1996ck,Smith:2005mm,Boyle:2005se}. The overall amplitude is determined by the Hubble scale during inflation and the spectrum has a cutoff at high frequency proportional to the reheating temperature~\cite{Nakayama:2008ip,Kuroyanagi:2008ye}. For the maximum reheating temperature of $10^{15}$\,GeV, for example, the cutoff frequency is $\sim 10^7$\,Hz. For the heavy enough inflaton, the highest frequency part can be dominated by the three-body decay contribution. If the inflaton is not heavy enough, the three-body decay contribution may be hidden by the inflationary background.

%%%%%%%%%%%%%%%%%%%%%%%%%%%%%%%%%%%%%%%%%%%%
\section*{Acknowledgments}
%%%%%%%%%%%%%%%%%%%%%%%%%%%%%%%%%%%%%%%%%%%%

YT would like to thank Yan-Qing Ma for helpful discussion. This work was supported by
the Grant-in-Aid for Scientific Research C (No.\ 18K03609 [KN]), and Innovative Areas (No.\ 16H06490 [YT],
No.\ 26104009 [KN], No.\ 15H05888 [KN], No.\ 17H06359 [KN]).

%%%%%%%%%%%%%%%%%%%%%%%%%%%%%%%%%%%%%%%%%%%%%%%%%%

%%%%%%%%%%%%%%%%%%%%%%%%%%%%%%%%%%%%%%%%%%%%%%%%%%

\end{document}